\documentclass[aps,prc,groupedaddress,showpacs,manuscript]{revtex4-1}
\usepackage{graphicx}

\usepackage{colordvi}

\begin{document}

\title{Collective flows of light particles in Au+Au collisions at intermediate energies}

\author {Yongjia Wang$\, ^{1,2}$,
Chenchen Guo$\, ^{2,3}$,
Qingfeng Li$\, ^{2}$\footnote{E-mail address: liqf@hutc.zj.cn},
Hongfei Zhang$\, ^{1}$,
Zhuxia Li$\, ^{4}$,
and
Wolfgang Trautmann$\, ^{5}$}

\affiliation{
1) School of Nuclear Science and Technology, Lanzhou University, Lanzhou 730000, P.R. China \\
2) School of Science, Huzhou Teachers College, Huzhou 313000, P.R. China \\
3) College of Nuclear Science and Technology, Beijing Normal University, Beijing 100875, P.R. China \\
4) China Institute of Atomic Energy, Beijing 102413, P.R. China\\
5) GSI Helmholtzzentrum f\"ur Schwerionenforschung GmbH, D-64291 Darmstadt, Germany\\
\\
 }
\date{\today}

\begin{abstract}
The Skyrme potential energy density functional is introduced into the
Ultrarelativistic Quantum Molecular Dynamics (UrQMD) model and
the updated version is applied to studying the
directed and elliptic flows of light particles (protons,
neutrons, deuterons, tritons, $^3$He and $^4$He) in $^{197}$Au+$^{197}$Au
collisions at beam energies 150, 250 and 400 MeV$/$nucleon. The
results are compared with the recent FOPI experimental data.
It is found that the yields and collective flows of light
particles can be described quite well. The influence of the equation
of state (EoS), medium-modified nucleon-nucleon elastic cross
sections (NNECS) and cluster recognition criteria on the directed
and elliptic flows is studied in detail. It is found that the
flows of light particles are sensitive to the medium-modified
NNECS, but not sensitive to the isospin dependent cluster
recognition criteria. It seems difficult, however, even with
the new data and calculations, to obtain a more accurate
constraint on the nuclear incompressibility $K_0$ than the
interval 200-260 MeV.
\end{abstract}



\maketitle
\section{Motivation}
The equation of state (EoS) of nuclear matter and the
nucleon-nucleon cross sections (NNCS) in the nuclear medium are hot topics in nuclear
physics since a long time~\cite{BALi08}. Heavy ion collisions (HICs) provide a unique
opportunity to study these subjects in the laboratories around the world. It has
been always difficult, however, to directly extract the information on the EoS and
NNCS from the measured quantities of HIC experiments because of the complexity of the
collision process and the restriction of the experimental data to the asymptotic
configurations recorded by the detectors. Microscopic transport theory has,
therefore, been a valuable tool for simulating the dynamical process of HICs, so as
to link the experimental observables to both the nuclear EoS and
the in-medium NNCS~\cite{Danie02}.

The collective flow is a common phenomenon of HICs,
first discovered at the Bevalac in 1984 (see Ref.~\cite{reisdorf1997} and
references therein). The directed flow (also named in-plane or sidewards flow)
and the elliptic flow
(also named out-of-plane flow) are two lower-order components of the flow which
have been widely used for studying HICs in a large range of beam energies varying
from tens of MeV up to several TeV per nucleon.
Newly measured experimental data of flows were usually compared with corresponding
theoretical results, calculated with the most recent updated theoretical transport
models, in order to obtain further insight into the properties of the EoS and
the in-medium NNCS. A large effort has been devoted
to constraining the stiffness of the EoS of isospin symmetric nuclear matter,
e.g., the pioneer works in Ref.~\cite{Hartnack:2005tr} with sub-threshold kaon production and in Ref.~\cite{Danie02} with collective flow observables,
with the result that it is most likely soft with an incompressibility $K_0$
of about 230 $\pm$ 30 MeV~\cite{Dutra:2012mb}. Up to now, however,
the stiffness of the EoS of isospin asymmetric nuclear matter (``$K_{asy}$''),
especially at high densities, is
still not well constrained and the medium modified NNCS have not been well
understood either. Thus both, more precise experimental data and self-consistent
theoretical models are still called for.

One of the interesting phenomena, already known from early-stage flow-related
experiments~\cite{Doss:1987kq,Gutbrod:1989wd,Gutbrod:1989gh}, is the dependence of
the directed and elliptic flows on the particle species. The flow effect is larger for
composite particles than for protons.
With the subsequent large number of experimental (see, e.g.,
Refs.~\cite{Westfall:1993zz,Partlan:1994vs,Huang:1996zz,Rami:1998bf,Barrette:1998bz})
and theoretical (see, e.g., Refs.~\cite{Peilert:1989kr,koch:1990,Zhang:1995zzb,Danielewicz:1994nb,Nemeth:1998by,Insolia:2000sr}) endeavors, the presence of this effect was confirmed by observing the increase of flow with the particle mass more precisely, even though the definitions of flow and the interpretations were somewhat different in the respective studies. Recently, by using the large acceptance apparatus FOPI at the Schwerionen-Synchrotron
(SIS) at GSI, a large amount of directed and elliptic flow data
for light charged particles (protons, deuterons, tritons, $^3$He and
$^4$He) from intermediate energy HICs have been made available
\cite{FOPI:2010aa,FOPI:2011aa}. Moreover, flows are presented differentially in the
FOPI data~\cite{FOPI:2011aa} in the form of both rapidity and transverse momentum
distributions. Therefore, new opportunities have been opened up which will allow us
to discuss the following questions:\\
(1) Is it possible to reduce the uncertainty of $K_0$ of the EoS by comparing a large
number of two-dimensional flow data with model calculations?\\
(2) Is it now possible to extract more information on the medium modifications of
NNCS? \\
(3) How do different cluster recognition criteria affect the flows of light particles?
This latter question arises because the newly developed isospin-dependent cluster
recognition method has been reported to affect the production of light
particles~\cite{Zhang:2012qm}.

The paper is arranged as follows. In the next section the new
version of the UrQMD transport model with the Skyrme potential
energy density functional is presented. In Section III, results
of collective flows of light particles from $^{197}$Au+$^{197}$Au
reactions at beam energies 150, 250 and 400 MeV$/$nucleon are shown.
Finally, a summary and outlook is given in Section IV.

\section{UrQMD model updates}
The UrQMD model~\cite{Bass98,Bleicher:1999xi,Li:2011zzp,Li:2012ta} has been widely and
successfully used to study \emph{pp}, \emph{p}A, and AA collisions within a large
energy range from Bevalac and SIS up to the AGS, SPS, RHIC, and LHC.
At lower energies, the UrQMD model is based on principles analogous
to the quantum molecular dynamics model (QMD) \cite{aichelin91} in which each
nucleon is represented by
a Gaussian wave packet in phase space. The centroids ${\bf r}_i$ and ${\bf p}_i$
of a nucleon $i$ in the coordinate and momentum spaces are
propagated according to Hamilton's equations of motion:

\begin{equation}
{\bf \dot{r}}_i=\frac{\partial H}{\partial {\bf p}_i},
\hspace{1.5cm} \mathrm{and} \hspace{1.5cm} {\bf
\dot{p}}_i=-\frac{\partial H}{\partial {\bf r}_i}. \label{Hemrp}
\end{equation}
The Hamiltonian $H$ consists of the kinetic energy $T$
and the effective two-body interaction potential energy $U$,
\begin{equation}
H=T+U
\end{equation}
with
\begin{equation}
T=\sum_i (E_i-m_i)=\sum_i (\sqrt{m_i^2+{\bf p}_i^2}-m_i), \label{HT}
\end{equation}
and
\begin{equation}
U=U_{\rho}+U_{md}+U_{coul}
\end{equation}
where $U_{coul}$ is the Coulomb energy, while the nuclear interaction potential energy terms $U_{\rho}$ and $U_{md}$ can be written as
\begin{equation}
 U_{\rho,md}=\int u_{\rho,md}d{\bf r} .
\end{equation}

In the current new version of the UrQMD model, the form of the momentum
dependent term $u_{md}$ is taken from the QMD model~\cite{aichelin91}
while the Skyrme potential energy density
functional $u_{\rho}$ is introduced in the same manner as in the improved
quantum molecular dynamics (ImQMD)
model~\cite{Zhang:2006vb,Zhang:2007gd} in which

\begin{eqnarray}
\label{equrho}
u_{\rho}=\frac{\alpha}{2}\frac{\rho^2}{\rho_0}+
\frac{\beta}{\eta+1}\frac{\rho^{\eta+1}}{\rho_0^{\eta}}+
\frac{g_{sur}}{2\rho_0}(\nabla\rho)^2
+\frac{g_{sur,iso}}{2\rho_0}[\nabla(\rho_n-\rho_p)]^2\nonumber\\
+(A\rho^{2}+B\rho^{\eta+1}+C\rho^{8/3})\delta^2
+g_{\rho\tau}\frac{\rho^{8/3}}{\rho_0^{5/3}}.
\end{eqnarray}
Here $\delta=(\rho_{n}-\rho_{p})/(\rho_{n}+\rho_{p})$ is the
isospin asymmetry defined through the neutron ($\rho_n$) and proton
($\rho_p$) densities with $\rho=\rho_n + \rho_p$.
The parameters $\alpha, \beta, \eta, g_{sur}$, and $g_{sur,iso}$ are related
to the Skyrme parameters via $\frac{\alpha}{2}=\frac{3}{8}t_{0}\rho_{0}$,
$\frac{\beta}{\eta+1}=\frac{1}{16}t_{3}\rho_{0}^{\eta}$,
$\frac{g_{sur}}{2}=\frac{1}{64}(9t_{1}-5t_{2}-4x_{2}t_{2})\rho_{0}$,
and $\frac{g_{sur,iso}}{2}=-\frac{1}{64}(3t_{1}(2x_{1}+1)+t_{2}(2x_{2}+1))\rho_{0}$.
The parameters $A$, $B$, and $C$ in the volume symmetry energy term of
Eq.~\ref{equrho} are
given by $A=-\frac{t_{0}}{4}(x_{0}+1/2)$, $B=-\frac{t_{3}}{24}(x_{3}+1/2)$,
and $C=-\frac{1}{24}(\frac{3\pi^{2}}{2})^{2/3}\Theta_{sym}$
where $\Theta_{sym}=3t_{1}x_{1}-t_{2}(4+5x_{2})$. The last term reads
$g_{\rho\tau}=\frac{3}{80}(3t_{1}+(5+4x_{2})t_{2})(\frac{3\pi^{2}}{2})^{2/3}\rho_{0}^{5/3}$.
The coefficients
$t_{0},t_{1},t_{2},t_{3}$ and $x_{0},x_{1},x_{2},x_{3}$ are the well-known parameters
of the Skyrme force.

In this work, we choose three sets of the Skyrme force, SkP
\cite{Dob84,Dutra:2012mb}, SV-mas08
\cite{Dutra:2012mb,Klupfel:2008af}, and SkA
\cite{Dutra:2012mb,Kohler} for incompressibility values $K_0$ varying
within 230 $\pm$ 30MeV. The main saturation properties of each
set are listed in Table \ref{skyrme} which shows that the saturation
density ($\rho_0$), the saturation energy
($E_0$), and the symmetry energy ($S_0$) at $\rho_0$ are close to
their commonly accepted values, $0.16$ fm$^{-3}$, $-16$ MeV, and
$32$ MeV, respectively. The other three parameters,
the slope $L$ of the symmetry energy, the symmetry incompressibility
$K_{asy}$, and the effective mass ratio $m^*/m$ at $\rho_0$, are also
found within their known regions of uncertainty.
It should be noticed that with the introduction of the ``standard" Skyrme potential energy density functional, these parameters are not varied independently. But, the effect of isovector part of EoS will not be much involved in this paper since its contribution to flows is much smaller than the isoscalar part of EoS.
\begin{table}[htbp]
\caption{\label{tab:table1} Saturation properties of three Skyrme parametrizations used in this work.}
\begin{ruledtabular}
\begin{tabular}{cccccccc}
 &SkP\cite{Dob84,Dutra:2012mb} & &SV-mas08\cite{Dutra:2012mb,Klupfel:2008af}& &SkA\cite{Dutra:2012mb,Kohler}\\
\hline

$\rho_{0}$ (fm$^{-3}$)& 0.163 && 0.160 && 0.155  \\
$E_0$ (MeV)        & -15.95 && -15.90 && -15.99 \\
$S(\rho_{0})(MeV)$ & 30.00 && 30.00 && 32.91 \\
$L (MeV)$          & 19.68 && 40.15 && 74.62 \\
$K_{asy}(MeV)$     & -266.60 && -172.38 && -78.46  \\
$m^{*}/m$          & 1.00 && 0.80 && 0.61  \\
$K_0(MeV)$         & 201 && 233 && 263  \\
\end{tabular}
\end{ruledtabular} \label{skyrme}
\end{table}

\begin{figure}[htbp]
\centering
\includegraphics[angle=0,width=0.9\textwidth]{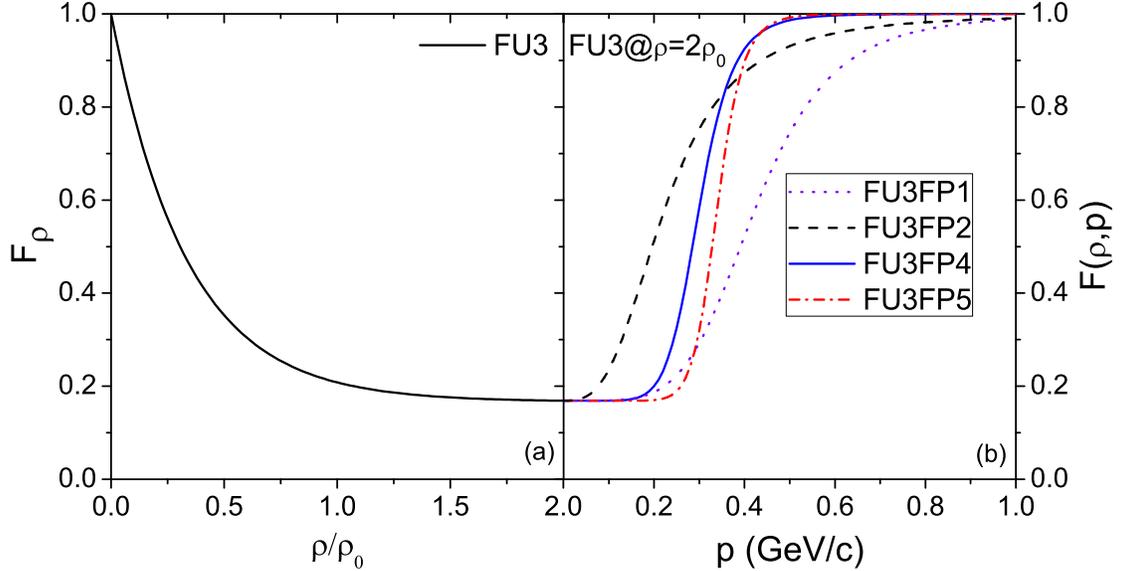}
\caption{\label{fig1}(Color online) (a) The medium correction factor $F_\rho$ obtained with the parameterization FU3 and (b)
the momentum dependence with the four options FP1, FP2, FP4, and FP5 given in Table \ref{tabfp} for FU3 at $\rho=2\rho_0$. }
\end{figure}

Concerning the NNCS, it is known that it will be modified by the nuclear
medium, according to approaches such as the (self-consistent) relativistic Boltzmann-Uehling-Uhlenbeck (RBUU) and the (Dirac-)Brueckner-Hartree-Fock (DBHF), which are based on the theory of quantum hydrodynamics (QHD), see e.g.,
Refs.~\cite{Chou:1984es,Li:1993rwa,Mao:1994zza,Li:2000sha,Sammarruca:2006jd}.
However, the details of this modification are still not clear. In this work,
as done previously~\cite{Li:2011zzp,Li:2006ez}, the in-medium nucleon-nucleon elastic
cross sections (NNECS) are treated to be factorized as the
product of a medium correction factor $F$ and the free cross sections.
For the inelastic channels, we still use the experimental free-space cross
sections which will not have a significant influence on results
studied in this work. The total nucleon-nucleon binary scattering
cross sections can thus be expressed as
\begin{equation}
\sigma_{tot}^{*}=\sigma_{in}+\sigma_{el}^{*}=\sigma_{in}+F(\rho,p)
\sigma_{el}  \label{ecsf}
\end{equation}
with
\begin{equation}
F(\rho,p)=\left\{
\begin{array}{l}
f_0 \hspace{3.45cm} p_{NN}>1 {\rm GeV}/c \\
\frac{F_{\rho} -f_0}{1+(p_{NN}/p_0)^\kappa}+f_0 \hspace{1cm}
p_{NN} \leq 1 {\rm GeV}/c
\end{array}
\right.
\label{fdpup}
\end{equation}
where $p_{NN}$ denotes
the relative momentum of two colliding nucleons. Here
$\sigma_{el}$ and $\sigma_{in}$ are the nucleon-nucleon elastic and
inelastic cross sections in free space, respectively, with the
proton-neutron cross sections being considered as different from the
proton-proton and neutron-neutron cross sections in accordance with
experimental data. The factor $F_\rho$ in Eq.\ \ref{fdpup} can be expressed as
\begin{equation}
F_\rho=\lambda+(1-\lambda)\exp[-\frac{\rho}{\zeta\rho_0}], \label{fr}
\end{equation}
which is also illustrated in Fig.\ \ref{fig1}(a). In this work, $\zeta$=1/3 and
$\lambda$=1/6 are adopted which corresponds to the
parametrization FU3 in Ref.~\cite{Li:2011zzp}.
The three parameters
$f_{0}$, $p_{0}$ and $\kappa$ in Eq.\ \ref{fdpup} can be varied in order
to obtain various momentum dependences of $F(\rho,p)$.

We select several parameter sets for this work which are shown in Table II. The
corresponding $F(\rho,p)$ functions at $\rho=2\rho_0$ are
illustrated in Fig.\ \ref{fig1}(b). The parameterizations FP1, FP2,
and FP3 were investigated and used in our previous
works \cite{Li:2011zzp,guocc,wangyj}. Specifically, the parameter set FU3FP1 was
used to investigate HICs around the balance energy where the
experimental data can be reproduced quite well with this set. Here,
we further introduce the FP4 and FP5 sets which lie
roughly between FP1 and FP2. This will permit more accurate tests of the
momentum dependence of the in-medium NNCS by taking advantage of the large
number of new FOPI data for directed and elliptic flows of light charged particles.
FP4 and FP5 differ mainly within
$p=0.2-0.4$ GeV/$c$ and the largest difference is within the narrow
region $p=0.25-0.35$ GeV/$c$. The treatment of the Pauli blocking
effect is the same as that in Ref.~\cite{Li:2011zzp}.

\begin{table}
\begin{center}
\renewcommand{\arraystretch}{1.2}
\begin{tabular}{|l|c|c|c|}\hline
\bf Set & $f_0$ & $p_0$ [GeV c$^{-1}$]   & $\kappa$  \\\hline\hline
\tt FP1 & 1   & 0.425 & \ 5  \        \\
\tt FP2 & 1   & 0.225 & 3        \\
\tt FP3 & 1   & 0.625 & 8        \\
\tt FP4 & 1   & 0.3 & 8        \\
\tt FP5 & 1   & 0.34 & 12        \\ \hline
\end{tabular}
\end{center}
\caption{The parameter sets FP1, FP2, FP3, FP4 and FP5 used for
describing the momentum dependence of $F(u,p)$. } \label{tabfp}
\end{table}

The UrQMD transport program stops at 250~fm$/c$ at which time a phase-space
coalescence mode~\cite{Kruse:1985pg} is used to construct clusters.
Usually, the minimum spanning tree (MST) algorithm is used. Recently,
an isospin-dependent MST (iso-MST)
method was introduced by Zhang {\it et al.}~\cite{Zhang:2012qm}.
Accordingly, in this work we will apply the two methods of fragment
recognition. The relative distance and momentum parameters $R_0$ and
$P_0$ are set to
$R_0^{nn}=R_0^{np}=R_0^{pp}=3.2$ fm for MST and
$R_0^{nn}=R_0^{np}=4.5$ fm and $R_0^{pp}=3.2$ fm for iso-MST and
$P_0 =0.25$ GeV/$c$ for both.

\section{Observables and Calculations}

\subsection{What to calculate}
Several hundred thousand events of $^{197}$Au+$^{197}$Au collisions for
each of the beam energies $E_{\text{lab}}$=150, 250, and 400 MeV$/$nucleon are
simulated randomly within the impact parameter region 0-7.5 fm, in order for small enough statistical error bars for observables. As
in Ref.~\cite{FOPI:2011aa}, the centrality is
characterized by the reduced impact parameter $b_0$ defined as
$b_0=b/b_{max}$, taking $b_{max} =1.15 (A_{P}^{1/3} +
A_{T}^{1/3})$~fm =13.4~fm for $^{197}$Au+$^{197}$Au.
At each beam energy, the calculations are
divided into 4 groups according to $b_0$:
$b_0<0.15$, $0.15<b_0<0.25$, $0.25<b_0<0.45$, and $0.45<b_0<0.55$ ($b_{max} \cdot 0.55 =7.4$~fm).
Five options of the UrQMD model differing in the treatment of the
mean-field potential (EoS), the medium modified NNCS and the
cluster recognition method are adopted and listed in Table
\ref{urqmd}. Clearly, the options UrQMD-I, UrQMD-IV, and UrQMD-V are for testing the influence of the mean
field potential, the options UrQMD-III and
UrQMD-IV are for testing the in-medium NNCS, and, UrQMD-II and UrQMD-IV are for testing the
influence of the cluster recognition method.
\begin{table}
\begin{center}
\renewcommand{\arraystretch}{1.2}
\begin{tabular}{|l|l|l|l|}\hline
\bf Set        &   EoS  &  NNCS            &  Cluster recognition  \\\hline\hline
\tt UrQMD-I    &   SkP       &  FU3FP4             &  iso-MST     \\
\tt UrQMD-II    &   SV-mas08  &  FU3FP4             &  MST      \\
\tt UrQMD-III   &   SV-mas08  &  FU3FP5             &  iso-MST       \\
\tt UrQMD-IV  &   SV-mas08  &  FU3FP4             &  iso-MST      \\
\tt UrQMD-V   &   SkA       &  FU3FP4             &  iso-MST      \\
\hline
\end{tabular}
\end{center}
\caption{Five options of the UrQMD transport model differing in
the treatments of the potential terms (EoS), of the
medium-modified NNCS, and of the cluster recognition method.}
\label{urqmd}
\end{table}

As a general test of the model, we first calculated fragment spectra as a function of
atomic number $Z$ for central $^{197}$Au+$^{197}$Au
collisions at beam energies $E_{\text lab}$=150, 250, and 400 MeV$/$nucleon.
It is found that results obtained by the five UrQMD options listed in
Table~\ref{urqmd} are in
agreement with experimental data and the difference among them are
relatively small. As a sensitive observable to both the EoS and the in-medium NNECS~\cite{FOPI:2010aa,Li:2011zzp}, the nuclear stopping quantity \emph{vartl}, defined by FOPI collaboration, of light charged clusters is investigated as well. One finds that results for flows and for the nuclear stopping follow in the same order when different treatments of the mean field and the collision terms are chosen. However, details on nuclear stopping calculations will be published elsewhere soon later. Since the aim of this work is to explore whether more
accurate constraints to the whole dynamic process of HICs can be obtained by
comparing with the new flow data of the FOPI collaboration,
we will not present results on the fragment spectrum and on stopping power in this paper.
  It is known that one of the most important observables to constrain the stiffness of EoS of nuclear
matter, especially at supra-normal densities, is the collective flow in HICs at
intermediate energies. Using the same parameterization as in
Ref.~\cite{FOPI:2011aa}, we have
\begin{equation}
 \frac{dN}{u_{t}du_{t}dyd\phi} = v_0 [1 + 2v_1 \cos(\phi) + 2v_2
\cos(2\phi)] ,
\end{equation}
in which the directed and elliptic flow parameters $v_1$ and $v_2$ can be written as:
  \begin{equation}
v_1\equiv \langle
cos(\phi)\rangle=\langle\frac{p_x}{p_t}\rangle;
v_2\equiv \langle cos(2\phi)\rangle=\langle\frac{p_x^2-p_y^2}{p_t^2}\rangle.
\label{eqv1}
\end{equation}
Here $\phi$ is the azimuthal angle of the emitted particle with
respect to the reaction plane, and $p_t=\sqrt{p_x^2+p_y^2}$ is the
transverse momentum of emitted particles. The angle brackets in
Eq.~\ref{eqv1} denote an average over all considered
particles from all events. The $v_{1}$ and $v_{2}$ have complex
multi-dimensional dependences. For a certain reaction with fixed
reaction system, beam energy, and impact parameter, they
are functions of $u_t$ and rapidity $y$. Here $u_t=\beta_t
\gamma$ is the transverse component of the four-velocity
$u$=($\gamma$, $\bf{\beta}\gamma$). We use the scaled units
$u_{t0}\equiv u_t/u_{1cm}$ and $y_0\equiv y/y_{1cm}$ as done
in \cite{FOPI:2011aa}, and the subscript $1cm$ denotes
the incident projectile in the center-of-mass system.

\begin{figure}[htbp]
\centering
\includegraphics[angle=0,width=0.9\textwidth]{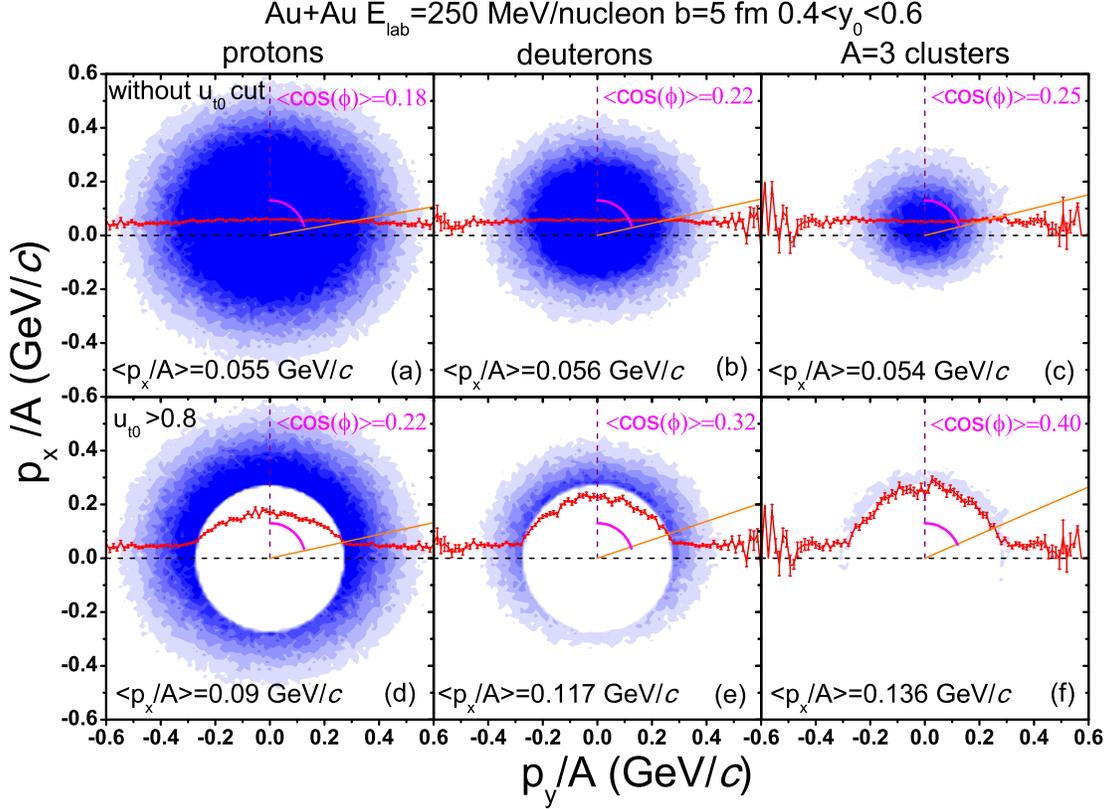}
\caption{\label{fig2} (Color online) Contour plots of $p_x/A$ vs
$p_y/A$ of free protons [left, (a) and (d)], deuterons [middle, (b) and (e)]
and $A=3$ clusters [right, (c) and (f)] calculated with UrQMD-IV and presented
without a $u_{t0}$ cut (upper row of panels) and with the cut $u_{t0}>0.8$
(lower panels) for $^{197}$Au+$^{197}$Au reactions at 250 MeV$/$nucleon,
$b=5$~fm, and the bin of forward rapidities $0.4<y_0<0.6$.
The solid lines represents the averaged $\langle{p_x}\rangle/{A}$ value for each
$\langle{p_y}\rangle/{A}$ bin while the displayed numerical values in the lower part of each panel are the
averages over the considered range of $\langle{p_y}\rangle/{A}$.
Values of $cos(\phi)$ are shown in the upper right corners of the panels.
}
\end{figure}

We first investigate how the condition $u_{t0}>0.8$, applied by FOPI to their
data, influences the
directed flow of different particles. In Fig.\ \ref{fig2}, the
$p_x/A$ vs. $p_y/A$ contour plots for emitted protons,
deuterons, and $A=3$ clusters (considering $^3H$ and $^3He$ results) are shown
without the $u_{t0}$ cut in the upper and with the cut $u_{t0}>0.8$ in the lower panels.
The interval of forward rapidities $0.4<y_0<0.6$ is selected,
so that more particles have positive $p_x$. The solid lines represent
the averaged $\langle{p_x}\rangle/{A}$ values for each $p_y/A$ bin and the
numerical values in the lower part of the panels are the averages over all
considered $p_y/A$ bins. In the upper right corner of each panel, the averaged value of $cos(\phi)$ ($=v_1$) for separate particles is also shown for comparison. It is apparent from the upper plots (a) -- (c)
that the $\langle{p_x}\rangle/{A}$ values of protons,
deuterons, and $A=3$ clusters are the same when the $u_{t0}$ cut
is not taken into account. When the cut $u_{t0}>0.8$ is applied, however,
shown in the lower plots (d) -- (f), the
$\langle{p_x}\rangle/{A}$ value increases with increasing particle mass.
If, however, the value of $v_1$ is examined, heavier clusters have larger transverse flow, even though there is no any $u_{t0}$ cut. With the consideration of the $u_{t0}$ cut, the effect of the particle species on flows becomes even more remarkable. This shows that the expected collective proportionality to the particle mass is observed when all particles are included and suggests that the phenomenon of an additional increase of the flow effect with the particle mass is strongly correlated with whether a transverse momentum cut is applied or not.

\begin{figure}[htbp]
\centering
\includegraphics[angle=0,width=0.9\textwidth]{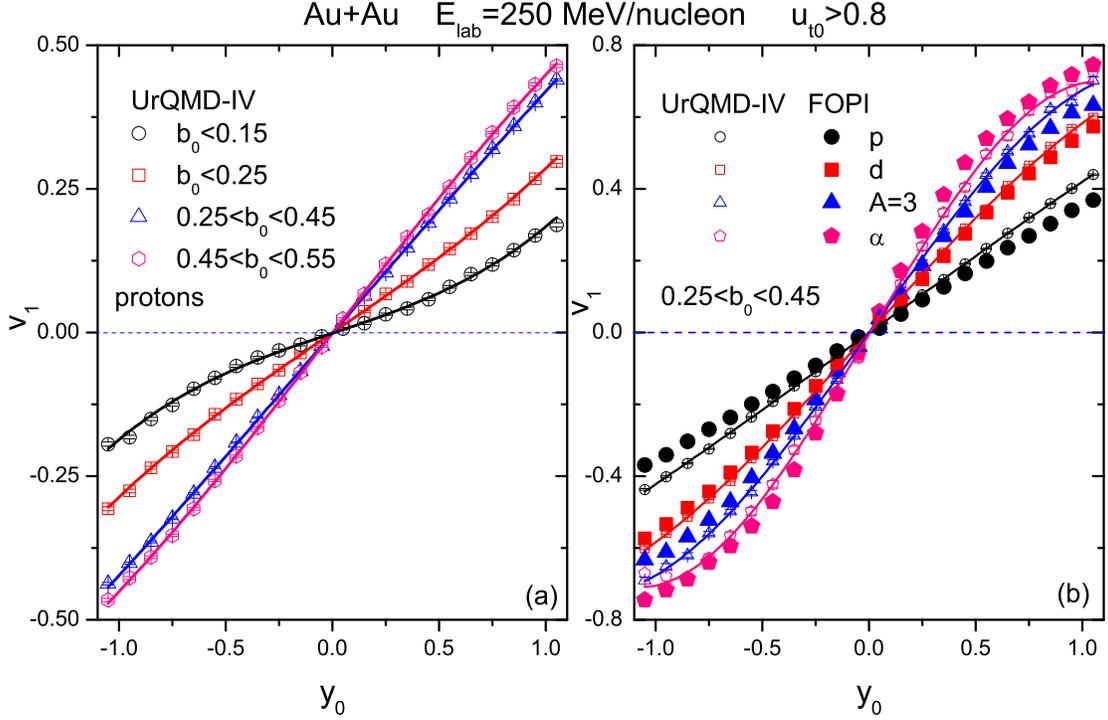}
\caption{\label{fig3}(Color online) Rapidity distribution of (a) the flow parameter
$v_1$ of protons under various centralities
and (b) flow parameter $v_1$
for protons, deuterons, $A=3$ clusters, and alpha particles
from $^{197}$Au+$^{197}$Au collisions at 250 MeV$/$nucleon with $0.25<b_0<0.45$,
as calculated with the
UrQMD-IV option (open symbols). The cut $u_{t0}>0.8$ is chosen.
The lines are fits to the calculation results (see text), while the
corresponding experimental data from Ref.\ \cite{FOPI:2011aa} are given by the
solid symbols.}
\end{figure}

Now, let us look at the collective flow as a function of rapidity when a $u_{t0}$ cut is applied.
Fig.\
\ref{fig3} shows the directed flow $v_1$ of protons under different centralities (open symbols) in plot (a), and $v_1$ of protons,
deuterons, $A=3$ and $\alpha$ particles (open symbols) with the centrality $0.25<b_0<0.45$ in (b) as a function of $y_0$.
 The UrQMD-IV is adopted for calculations, the reaction conditions in Fig\
\ref{fig3}(b) are chosen to be the same as the FOPI experimental data (solid symbols) of Ref.~\cite{FOPI:2011aa}.
The solid curves in the figure are fits to calculation
results assuming $v_1(y_0)=v_{11}\cdot y_0 + v_{13}\cdot y_0^3+c$
in the range of $-1.1<y_0<1.1$. The fit also provides the
slope value $v_{11}$ of $v_1$ at $y_0=0$ which will be discussed later.
In Fig.~\ref{fig3}(b), it is found that our calculated results for all particles considered are in agreement
with the experimental data in the whole rapidity region.

\begin{figure}[htbp]
\centering
\includegraphics[angle=0,width=0.9\textwidth]{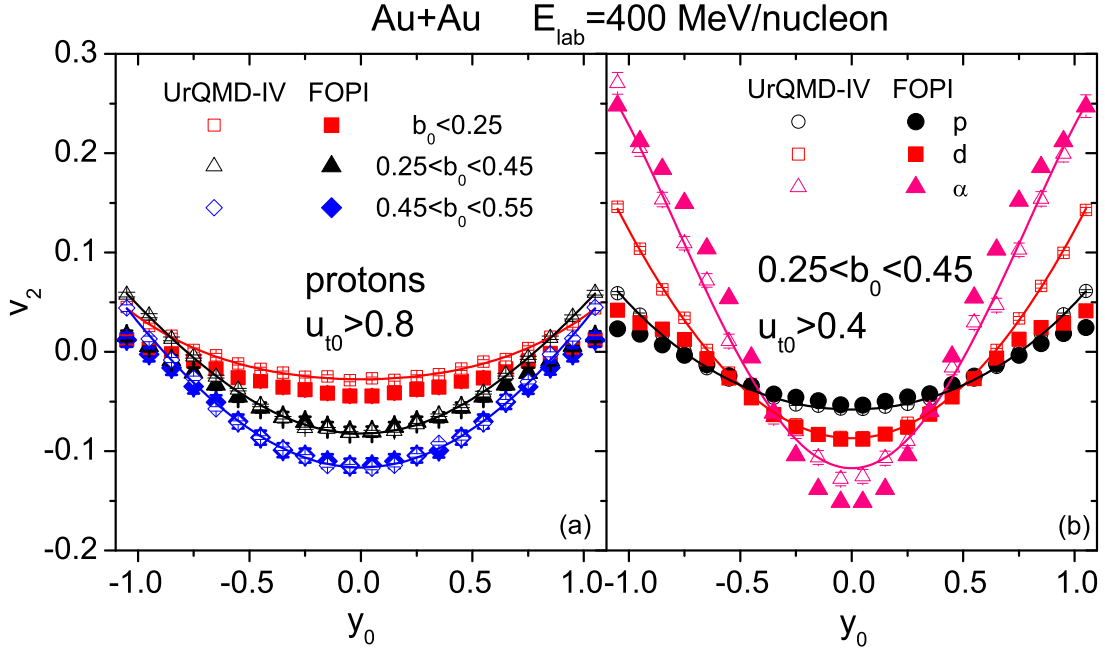}
\caption{\label{fig4} (Color online) Rapidity distribution of the flow parameter
$v_2$ of protons for various centralities (a) and the $v_2$ of
protons, deuterons, and alpha particles for the impact-parameter
bin $0.25<b_0<0.45$ for $^{197}$Au+$^{197}$Au
collisions at 400 MeV$/$nucleon. Calculations
with UrQMD-IV are shown with open symbols while the FOPI data,
taken from Ref.\ \cite{FOPI:2011aa}, are shown by solid symbols.
The lines are fits to the calculated results (see text).}
\end{figure}

The elliptic flow $v_2$ of light particles is also calculated and compared with
FOPI data from Ref.\ \cite{FOPI:2011aa}. In Fig.\ \ref{fig4}, the results of
calculations with UrQMD-IV and the FOPI data from $^{197}$Au+$^{197}$Au collisions
at 400 MeV$/$nucleon are represented by the open and solid symbols, respectively.
In the left panel, the elliptic flow parameter $v_2$ of protons as a function of
$y_0$ is shown for three centralities, while the $v_2$ for different particles,
i.e., protons, deuterons and alpha particles, is given in the right panel
(for semi-central collisions and with the less restrictive $u_{t0}$ cut applied by FOPI
at the higher energy).
The figure shows that the FOPI $v_2$ flow data, within a large centrality region
and for several particles, can also be quite well described with the updated
UrQMD transport model.
Further, with the fit $v_2(y_0)=v_{20} + v_{22}\cdot y_0^2 +
v_{24}\cdot y_0^4$ to the calculation, the elliptic flow at mid-rapidity, $v_{20}$,
can be obtained.
\subsection{Effects of EoS, NNCS, and cluster recognition on flows of light particles}

\begin{figure}[htbp]
\centering
\includegraphics[angle=0,width=0.9\textwidth]{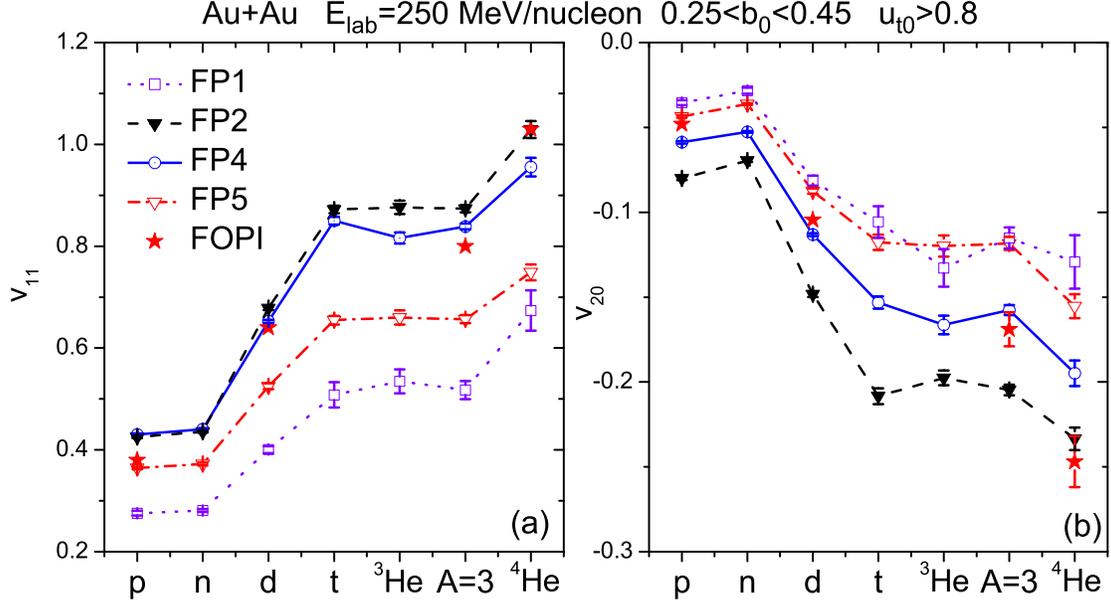}
\caption{\label{fig5} (Color online) (a) The $v_{11}$ and (b) the $v_{20}$  values for
light particles up to mass number $A=4$ calculated with FP1, FP2, FP4 and FP5 (lines with symbols) while other inputs are the same as those in the UrQMD-IV set.
The reaction $^{197}$Au+$^{197}$Au
at the beam energy 250 MeV$/$nucleon with $0.25<b_0<0.45$ is considered as an example. The FOPI
experimental data (stars) are taken from Ref.\ \cite{FOPI:2011aa}.}
\end{figure}

\begin{figure}[htbp]
\centering
\includegraphics[angle=0,width=0.7\textwidth]{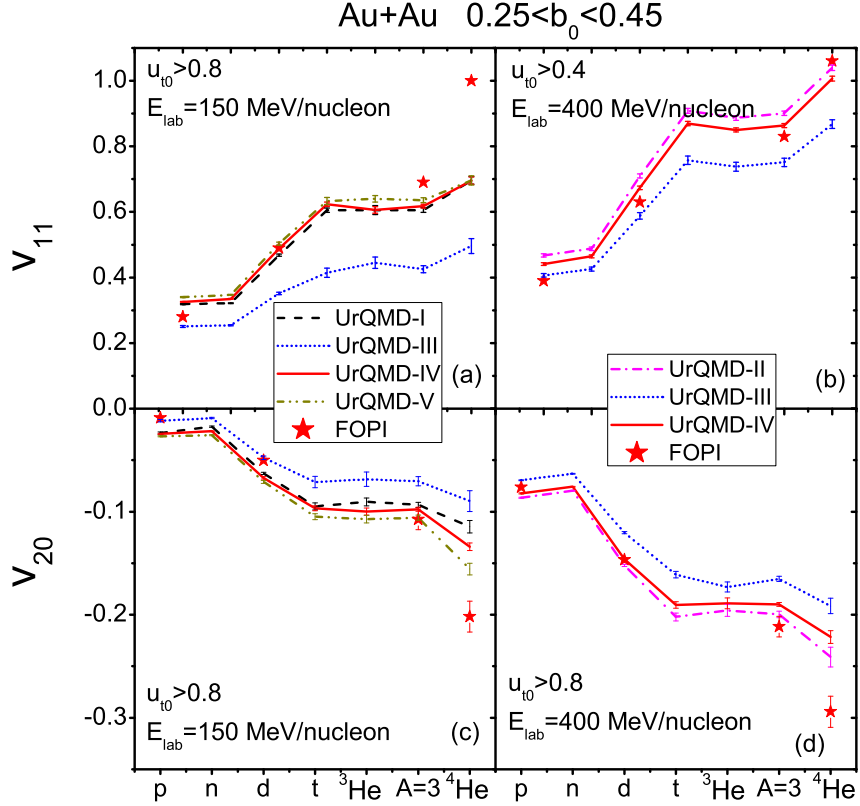}
\caption{\label{fig6} (Color online) $v_{11}$ [(a) and (b)]
and $v_{20}$ [(c) and (d)] for light particles from
semi-central ($0.25<b_0<0.45$) $^{197}$Au+$^{197}$Au collisions
at $E_{\text lab}=150$ (left) and 400 MeV$/$nucleon (right). The calculations
performed with five UrQMD parameter sets are
distinguished by different lines as indicated. The FOPI experimental data
from Ref.\ \cite{FOPI:2011aa} are shown by stars. }
\end{figure}

In order to show why the sets FP4 and FP5 have been introduced in addition
to FP1 and FP2 used previously for testing the
momentum dependence of the in-medium NNCS, we display in Fig.\ \ref{fig5}(a)
the $v_{11}$
and in Fig.\ \ref{fig5}(b) the $v_{20}$ values for
light particles calculated with the four sets FP1, FP2, FP4, and FP5.
Other inputs are the same as those in the UrQMD-IV set. Firstly, one sees clearly that
calculation results with FP4 and FP5 are well separated. It means
that the directed and elliptic flows of light
particles are very sensitive to the exact momentum dependence of in-medium
NNCS within a narrow region of $p=0.2-0.4$
GeV/$c$, which is due to a larger number of collisions happened in such a relative momentum region. Secondly, the $v_{11}$ of light particles calculated with FP2 and FP4 and
the $v_{20}$ calculated with FP1 and FP5
are very close to each other, respectively. Remembering that there is a large
difference between FP2 and FP4 at the low momentum part and
between FP1 and FP5 at high momenta (see Fig.\ \ref{fig1}),
we may conclude that the directed flow of light particles is not sensitive to the
low momentum part while the elliptic flow is not sensitive to the high
momentum part of the momentum dependent NNECS. However, with the further increase of beam energy, the sensitivity of the collective flow to the parameterization FP4 and FP5 will be reduced since they overlap at higher relative momentum (which is also shown in Fig.\ \ref{fig6}). The figure finally also shows
that the calculations with FP4 can best reproduce the experimental data.

\begin{figure}[htbp]
\centering
\includegraphics[angle=0,width=0.9\textwidth]{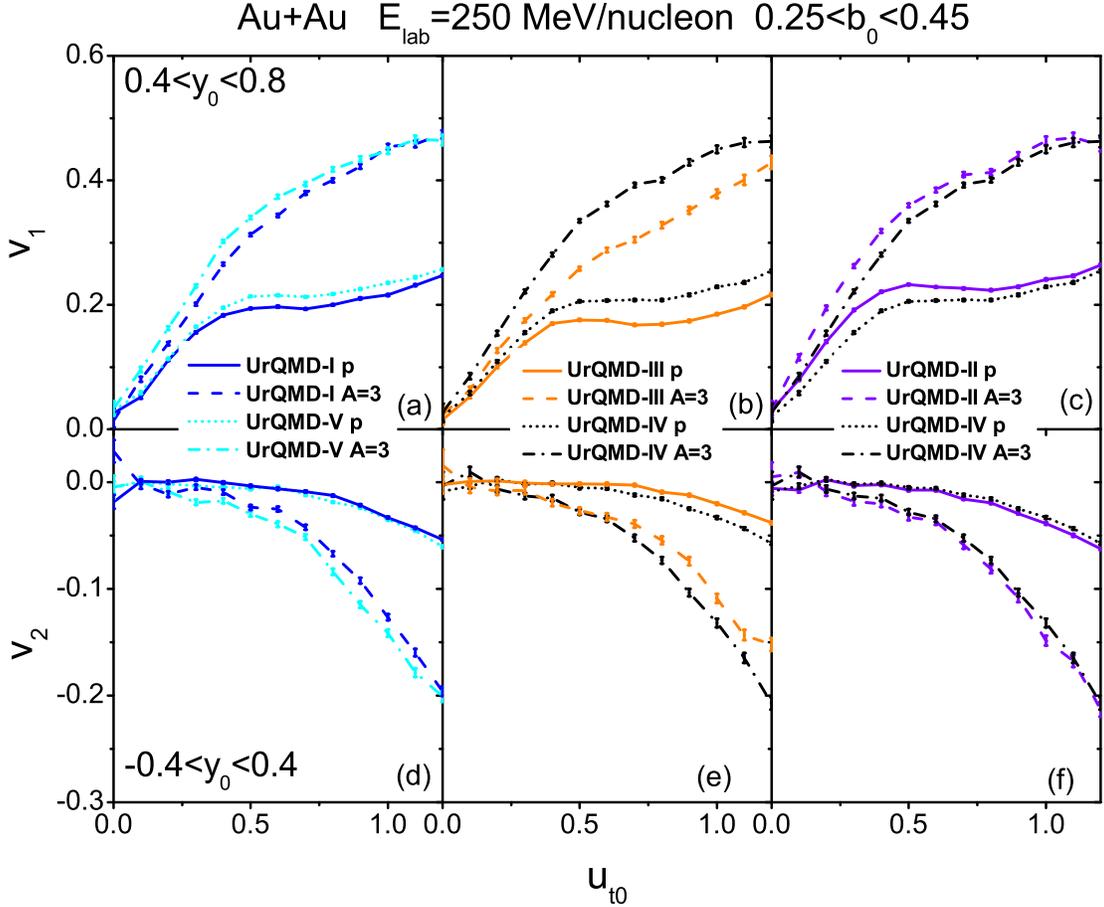}
\caption{\label{fig7} (Color online) Parameters $v_1$ of directed flow (upper panels)
and $v_2$ of elliptic flow (lower panels) for protons and $A=3$ clusters as
a function of $u_{t0}$. Calculations are obtained with UrQMD-I and UrQMD-V in (a)
and (d), UrQMD-III and UrQMD-IV in (b) and (e), and UrQMD-II and UrQMD-IV in (c)
and (f), respectively. The reaction $^{197}$Au+$^{197}$Au at the beam energy 250 MeV$/$nucleon with $0.25<b_0<0.45$
is considered as an example. The rapidity cuts $0.4<y_0<0.8$ and $|y_0|<0.4$ are
chosen for $v_1$ and $v_2$, respectively. }
\end{figure}

Besides the medium modification on NNECS, also the influence of the mean field
and of the cluster recognition method on flows is further investigated.
In Fig.\ \ref{fig6}, the $v_{11}$ and $v_{20}$ values obtained from calculations
with different UrQMD sets for light particles from
semi-central $^{197}$Au+$^{197}$Au collisions
at two beam energies, 150 (left) and 400 MeV$/$nucleon (right),
are compared with the FOPI data. 
More specifically, the mean field effect is examined in Fig.\
\ref{fig6}(a) and (c) where calculations with UrQMD-I, UrQMD-III, UrQMD-IV and
UrQMD-V sets are shown, while the cluster recognition effect is tested in
Fig.\ \ref{fig6}(b) and (d) with calculations using the UrQMD-II, UrQMD-III
and UrQMD-IV sets.
One immediately sees that, for both $v_{11}$ and $v_{20}$,
calculations with UrQMD-I, UrQMD-II, UrQMD-IV and UrQMD-V are grouped together,
while absolute values obtained with UrQMD-III are apparently
smaller, especially  for composite particles. The main reason
is that, with FP5, the reduction of the in-medium cross section is stronger
in the UrQMD-III case. Flows of composite particles at intermediate
energies are, apparently, very useful to test the behavior of
the momentum dependence of in-medium NNECS, especially in the momentum region
$p$=0.2-0.4 GeV$/c$. Secondly, although the absolute values of $v_{11}$ and $v_{20}$ are still seen to
increase gradually with the increasing incompressibility $K_0$ of the EoS, by
examining calculations going from UrQMD-I, UrQMD-IV, to UrQMD-V sets,
the differences between them are too small to extract a more
accurate $K_0$ value than 230 $\pm$ 30 MeV from the present calculations
and experimental data. By employing a much stiffer EoS such as SIII (with $K_0$=355 MeV), we have checked that the sensitivity of the flows to the EoS is comparable with previous studies shown, e.g., Ref.~\cite{Danie02}. The insensitivity of the flows to the EoS shown here is only due that the selected range of $K_0$ values is rather narrow based on the latest progress on it. It is further noticed that, although the effective mass values of the three Skyrme forces are largely different, the flow is not influenced significantly. It is because that the momentum dependent terms in the Skyrme potential energy density functional are obtained by the Thomas-Fermi approximation to the kinetic energy density and can not fully represent the momentum dependence of the whole non-equilibrium dynamic process. We note that, in order
to obtain an improved flow data set of light fragments for $^{197}$Au+$^{197}$Au
collisions and to extend the study of the density dependent symmetry energy to
other systems,
a new experiment (S394) was recently carried out at the GSI laboratory by the
ASY-EOS collaboration \cite{Russotto:2012jb}. It is certainly hopeful for us to
further reduce the uncertainties in both $K_0$ and $K_{asy}$ with the help of the
new experiment. Finally, from Fig.\ \ref{fig6}(b) and (d) one finds that,
no matter which flow parameter is chosen, the difference between results calculated
with UrQMD-II and with UrQMD-IV is also very small. And, the flow parameter difference in some isospin partners such as proton+neutron and $^{3}He$+$^{3}H$ is not obvious as well. It indicates that the different
cluster recognition methods MST and iso-MST have only a weak
effect on the flow parameters. However, the effect should depend on the transverse momentum and rapidity cuts (as will be seen in Fig.\ \ref{fig7}(c), the difference in $v_1$ between MST and iso-MST becomes larger as $u_{t0}$ decreases). Since both MST and iso-MST are different treatments related to isospin in the coalescence model at freeze-out, it
 has been found in Ref.~\cite{Zhang:2012qm} that yields of neutron-rich lighter fragments as well as isospin-dependent observables such as yield ratios between isospin partners are influenced. Similarly, the flow difference or ratio between isospin partners in some transverse momentum and/or rapidity windows might be influenced by the consideration of isospin in MST, which deserves further investigation.
Generally speaking,
we can conclude that
the new FOPI flow data can be reproduced by the UrQMD model calculations when
the FU3FP4 medium modification of NNECS is adopted, with the
only exception of $\alpha$ particle flow which is underestimated. Reasons for the underestimation will be discussed in the next subsection.

In order to see more clearly effects of the mean field
potential, the in-medium NNCS, and the cluster recognition method
on flows, the calculated parameters $v_1$ of directed and $v_2$ of elliptic flow
are shown as a function of $u_{t0}$ in Fig.\ \ref{fig7}.
For this purpose, we compare results of
protons and $A=3$ clusters obtained
with UrQMD-I and UrQMD-V in (a) and (d), with UrQMD-III and
UrQMD-IV in (b) and (e), and with UrQMD-II and UrQMD-IV in (c) and (f),
respectively. As an example, $^{197}$Au+$^{197}$Au collisions at the beam
energy 250 MeV$/$nucleon with $0.25<b_0<0.45$ are chosen. One sees a significant
effect on both flow parameters only in the case of the comparison
of calculations with UrQMD-III to UrQMD-IV shown in (b) and (e), especially for $A=3$
clusters.
This situation is quite similar to that shown in Fig.\ \ref{fig6}. Further, it can be seen from Fig.\ \ref{fig7}
that at about $u_{t0}>0.5$
the effect of medium modified NNCS on flows of $A=3$ clusters is enlarged [from (b) and (e)] while the
other two effects are reduced so that one may be able to
more cleanly determine the medium modifications of NNCS in this momentum region.
\begin{figure}[htbp]
\centering
\includegraphics[angle=0,width=0.9\textwidth]{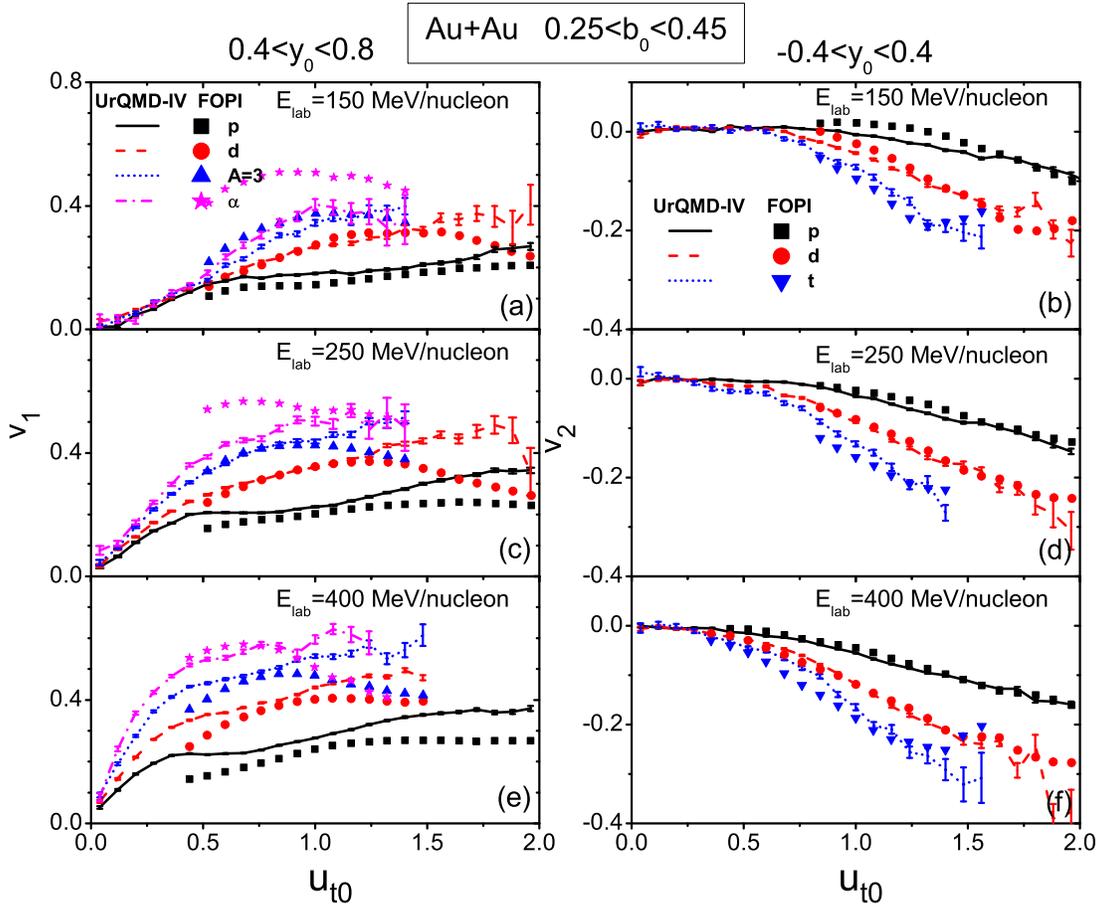}
\caption{\label{fig8} (Color online) The $u_{t0}$ dependence of parameters $v_1$ of
directed (left) and $v_2$ of elliptic flows (right) of light charged particles from
semi-central ($0.25<b_0<0.45$) $^{197}$Au+$^{197}$Au collisions at beam energies
150 [(a) and (b)], 250 [(c) and (d)], and 400
MeV$/$nucleon [(e) and (f)]. The rapidity cuts $0.4<y_0<0.8$ and $|y_0|<0.4$ are
chosen for $v_1$ and $v_2$, respectively.
Calculated results
with UrQMD-IV are represented by different lines as indicated, the FOPI experimental
data from Ref.\ \cite{FOPI:2011aa} are shown by solid symbols. }
\end{figure}

\subsection{Comparison of calculated $u_{t0}$ dependent flows to experimental data}
We finally show in Fig.\ \ref{fig8} the $u_{t0}$ dependence of calculated directed
(left panels) and elliptic flows (right panels) of light charged particles at beam
energies 150, 250, and 400 MeV$/$nucleon (lines).
The reaction system and chosen rapidity cuts are the same as for the
experimental data taken from Ref.\ \cite{FOPI:2011aa} and shown by the full symbols.
It is firstly observed that calculations with the UrQMD-IV set
reproduce the $v_1$ and $v_2$
data reasonably well with some exceptions. Although the experimental data of directed
flow of $\alpha$ particles can not be well described by the
model, the relatively large flow effect is clearly exhibited in plots (a), (c) and (e)
of Fig.\ \ref{fig8}.
Secondly, calculation results for absolute $v_1$ and $v_2$ values of protons are
slightly larger than the FOPI data, which is similar to the simulation
results shown in Ref.\ \cite{FOPI:2011aa} where the IQMD model was used.
Thirdly, when $u_{t0}$ is larger than about 1.0, the
deviation of the calculated $v_1$ from the data starts to increase in some of the
particle cases.
Although, on the other hand, the yields of these particles are quite small in these
$u_{t0}$ and $y_0$ regions and the contribution to the final $v_{11}$ value is
thus very limited. One has indeed seen the successful description of the
$u_{t0}$-integrated data by
the UrQMD-IV set shown in Fig.\ \ref{fig6}.

To our knowledge, the discrepancies shown above can be (partly) understood since (1), it is argued that, largely due to simplifications in the initial wave function of particles (nucleons and possible clusters) and quantum effects in two-body collisions, the yield of free nucleons (intermediate mass fragments, especially alpha particles), is largely overestimated (underestimated) by QMD. And, because of the strong decay of excited fragments, flows of lighter particles will definitely inherit partly those of their heavier parent fragments. Therefore, some of the free nucleons might thus actually belong to fragments. Since the flow effect is larger for fragments than for emitted nucleons, the calculated flows of free protons are consequently overestimated. As for alpha particle, however, the calculated flows are larger underestimated especially at small $u_{t0}$ for HICs at lower beam energies [as can be seen in Fig.\ \ref{fig8}(a)], which might be due to its deficiency of production from heavier excited fragments and its instability after production in model calculations. In order to make it better, the antisymmetrized molecular dynamics model (AMD) has been around and being kept updating~\cite{ono,KanadaEn'yo:2012bj}. (2), the importance of the optical potential to observables such as particle production and flow measured in HICs at intermediate energies has been widely investigated but its form is still far from settled~\cite{aich3,hart4,chen5}. And (3), the treatment of the fragment production could be more comprehensive than the current constraints in the phase space besides the consideration of the isospin. For example, in order to describe the early formation of fragments, the simulated annealing clusterization algorithm (SACA)~\cite{Puri:1996qv}, which is based on the energy minimization criteria, was proposed and shown promising.

Finally, as for flows of deuterons and $A=3$ clusters, it is seen that the comparison
of UrQMD-IV calculations with the experimental data is fairly good in the range
$0.5<u_{t0}<1.0$. In view of the result shown in Fig.\ \ref{fig7}, it is highly
advantageous to investigate the detailed behavior of the medium corrected NNCS
in this momentum region.
In order for a more reliable comparison to data, some recently concerned issues in the community, i.e., the internal magnetic fields \cite{Ou:2011fm} and non-central forces as, e.g., the tensor force and spin-orbit coupling \cite{Xu:2009bb,Xu:2012hh} which might influence the freeze-out mode of HICs especially for non-central collisions at large momenta and rapidities, will be further studied within the same transport theory.

\section{Summary and Outlook}
In summary, we have studied the directed and elliptic flows of
light particles in $^{197}$Au+$^{197}$Au collisions at beam energies 150, 250 and 400
MeV$/$nucleon by using the updated UrQMD model in which the Skyrme
potential energy density functional is introduced. After the detailed study of the
influence of equation of state (EoS), medium-modified
nucleon-nucleon elastic cross section (NNECS) and cluster
recognition criteria on flows, the three
questions asked in the introduction can be answered: (1) it is difficult to get
a more exact value of the incompressibility from the present flow data
than $K_0$ =230 $\pm$ 30~MeV, (2) the different choices of medium-modified NNECS
exhibit a significant influence on the light particle flows and,
particularly, on the flows of light composite particles;
(3) the influence of the cluster recognition method on cluster flows is weak. The
version of UrQMD-IV, comprising the SV-mas08 force with a corresponding
incompressibility $K_0$=234~MeV, the FU3FP4 medium-modified NNECS and the iso-MST
cluster recognition method, describes the directed and elliptic flows of light
particles as functions of both rapidity and transverse momentum rather well.

Theoretically, the spin-orbit coupling term in the Skyrme interactions will be
further put into the UrQMD transport model after incorporating the spin degree of
freedom and its contribution
to flows, especially at large rapidities and/or transverse momenta, for intermediate
energy HICs can then be identified.
Together with the forthcoming new flow data of light particles measured by the ASY-EOS
collaboration at GSI, we hope to further reduce the uncertainties in both $K_0$
and $K_{asy}$ of the isospin-dependent EoS within the present framework of UrQMD in
the near future.

\begin{acknowledgements}
We acknowledge support by the computing server C3S2 in Huzhou
Teachers College. The work is supported in part by the National
Natural Science Foundation of China (Nos. 10905021, 10979023,
11175074, 11075215, 11275052, 11375062), the Zhejiang Provincial Natural
Science Foundation of China (No. Y6090210), the Qian-Jiang
Talents Project of Zhejiang Province (No. 2010R10102), and the National
Key Basic Research Program of China (No. 2013CB834400).
\end{acknowledgements}

\end{document}